\documentstyle[12pt,epsfig]{article}

\def    \bc            {\begin{center}}
\def    \ec            {\end{center}}

\def    \dd            {\displaystyle}
\def    \be            {\begin{equation}}
\def    \ee            {\end{equation}}
\def    \bea           {\begin{eqnarray}}
\def    \eea           {\end{eqnarray}}
\def    \nn            {\nonumber}

\def    \Ag            {\mbox{$ A_{\lambda{\bar\lambda}}^\gamma $}}
\def    \Az            {\mbox{$ A_{\lambda{\bar\lambda}}^Z $}}
\def    \dAg           {\mbox{$ \delta A_{\lambda{\bar\lambda}}^\gamma $}}
\def    \dAz           {\mbox{$ \delta A_{\lambda{\bar\lambda}}^Z $}}
\def    \dAV           {\mbox{$ \delta A_{\lambda{\bar\lambda}}^V $}}

\def    \M2            {\mbox{$ \tilde{\cal M} $}}
\def    \Mg            {\mbox{$ {\tilde{\cal M}}^\gamma $}}
\def    \Mz            {\mbox{$ {\tilde{\cal M}}^Z $}}
\def    \Mnu           {\mbox{$ {\tilde{\cal M}}^\nu $}}
\def    \thetab        {\bar\theta}
\def    \raw           {\rightarrow}

\catcode`@=11
\def\marginnote#1{}
\newcount\hour
\newcount\minute
\newtoks\amorpm
\hour=\time\divide\hour by60
\minute=\time{\multiply\hour by60 \global\advance\minute by-\hour}
\edef\standardtime{{\ifnum\hour<12 \global\amorpm={am}%
        \else\global\amorpm={pm}\advance\hour by-12 \fi
        \ifnum\hour=0 \hour=12 \fi
        \number\hour:\ifnum\minute<10 0\fi\number\minute\the\amorpm}}
\edef\militarytime{\number\hour:\ifnum\minute<10 0\fi\number\minute}
\def\draftlabel#1{{\@bsphack\if@filesw {\let\thepage\relax
   \xdef\@gtempa{\write\@auxout{\string
      \newlabel{#1}{{\@currentlabel}{\thepage}}}}}\@gtempa
   \if@nobreak \ifvmode\nobreak\fi\fi\fi\@esphack}
        \gdef\@eqnlabel{#1}}
\def\@eqnlabel{}
\def\@vacuum{}
\def\draftmarginnote#1{\marginpar{\raggedright\scriptsize\tt#1}}
\def\draft{\oddsidemargin 0.0truein
        \def\@oddfoot{\sl preliminary draft \hfil
        \rm\thepage\hfil\sl\today\quad\militarytime}
        \let\@evenfoot\@oddfoot \overfullrule 3pt
        \let\label=\draftlabel
        \let\marginnote=\draftmarginnote
   \def\@eqnnum{(\theequation)\rlap{\kern\marginparsep\tt\@eqnlabel}%
\global\let\@eqnlabel\@vacuum}  }
\catcode`@=12
\begin{document}
\renewcommand{\thefootnote}{\alph{footnote}}
\begin{titlepage}
\vspace*{-1cm}
%\phantom{bla}
%\hfill{DFPD/96/66}
%\\
%\phantom{bla}
%\hfill{DFPD~95/TH/66}
%\\
\phantom{bla}
\hfill{hep-ph/9611414}
\vskip 2.5cm
\begin{center}
{\Large\bf Sum rules for asymptotic form factors 
\\
\vskip .4cm
in $e^+e^- \to W^+W^-$ scattering}
%\footnote{Work supported in part by the European Union under
%contract No.~CHRX-CT92-0004.}
\end{center}
\vskip 2.0cm
\begin{center}
{\large F. Feruglio \footnote{feruglio@padova.infn.it}
and S. Rigolin \footnote{rigolin@padova.infn.it}}
\\
\vskip .1cm
{\it Dipartimento di Fisica, Universit\`a di Padova and INFN, I-35131 Padua, Italy}
%\\
%\vskip .2cm
%and
%\\
%\vskip .2cm
%{\large Author n+1, Author n+2,...}
%\\
%\vskip .1cm
%Dipartimento di Fisica, Universit\`a di Padova, I-35131 Padua, Italy
%\\
%\vskip .1cm
%Theory Division, CERN, CH-1211 Geneva 23, Switzerland
\end{center}
\vskip 2.0cm
\begin{abstract}
\noindent
At very large energies and in $SU(2)_L\otimes U(1)_Y$ gauge 
theories, the trilinear gauge boson vertices relevant for
$e^+e^- \to W^+W^-$ scattering are related in a simple way 
to the gauge boson self-energies. We derive these relations, 
both from the requirement of perturbative unitarity and from
the Ward identities of the theory. 
Our discussion shows that, in general, it is never possible to neglect
vector boson self-energies when computing the form factors
that parametrize the $e^+ e^- \to W^+ W^-$ helicity amplitudes.
The exclusion of the self-energy contributions would lead to
estimates of the effects wrong by orders
of magnitude.
We propose a simple way of including the self-energy contributions 
in an appropriate definition of the form factors. 
%As a consequence
%of these relations we show how, neglecting the self-energy
%contributions in the evaluation of the form factors
%parametrizing $e^+e^- \to W^+W^-$, may generally
%result in wrong estimates. 
\end{abstract}
\vfill{
DFPD 96/TH/66
\newline
\noindent
November 1996}
\end{titlepage}
\setcounter{footnote}{0}
\renewcommand{\thefootnote}{\arabic{footnote}}
\vskip2truecm
\vspace{1cm}
%Role of $e e \to W W$ process in present and future  e e machines:
%physical goals.
%\vskip .5cm
{\bf 1.}
It has become customary to analyse the $W$ pair production in $e^+ e^-$
machines by parametrizing the scattering amplitude in terms of
a set of form factors characterizing the most general trilinear
gauge vertex (TGV) involving a neutral vector boson $V$ ($V=\gamma,Z$)
and two on-shell $W$'s \cite{gae,hpz}. By studying the angular distributions
of the W it is possible to extract some
informations on the form factors, thus providing new tests of the
underlying electroweak theory \cite{hpz,bk}.
Any SM extension can be easily constrained, by analysing
directly the contribution of the new particles and/or interactions 
to the different anomalous couplings.
Indeed, in any recent analysis of $e^+ e^- \to
W^+ W^-$ \cite{boo,arg}, the above mentioned form factors play a crucial role
and represent an important ingredient for a meaningful comparison
between theory and experiment. For this reason any new information
on the trilinear gauge vertices is welcome, particularly if
it comes from the theory. 

Purpose of this note is to show that, at large energies, the TGV relevant 
to the 
$e^+ e^- \to W^+ W^-$ process satisfy a set of relations that will be 
called sum rules. 
These relations can be viewed as a direct consequence of perturbative 
unitarity, that is the physical requirement that, at a given loop-order, 
the bad unitarity behaviour, which potentially affects the scattering 
amplitudes in massive Yang-Mills theories, disappears thanks to a
cancellation among the dangerous contributions. Indeed, in the first part
of this note we derive the sum rules by discussing the one-loop
renormalization of the $e^+ e^- \to W^+ W^-$ scattering, and by
imposing unitarity to the amplitude obtained.

It has been known since a long time that there is a deep relation
between unitarity and gauge invariance \cite{clt} and in the second part of 
this
work we re-derive the sum rules starting from the Ward identities 
of a spontaneously broken $SU(2)\otimes U(1)$ theory.

Although the sum rules presented here are valid only in the limit
of very large energies, we will see how they qualitatively constrain
also the region of experimental interest, namely the one between
the electroweak scale $m_W$ and the scale $M$ associated to new physics.
An important feature of the sum rules is that they involve combinations
of the three-point and the two-point functions among vector bosons.
This emphasizes the importance of including the vector boson self-energies 
for a correct evaluation of the form factors which
parametrize the $e^+ e^- \to W^+ W^-$ process. 
Use of form factors explicitly involving 
only the TGV may easily lead
to wrong conclusion about the corresponding cross-section.
%unless the form factors are appropriately re-defined to contain
%the contribution of the self-energies. 
We will discuss these issues in a concluding paragraph.
\vskip .5cm

{\bf 2.}
We start the discussion by summarizing the standard parametrization
of the $e^+e^- \to W^+W^-$ helicity amplitudes \cite{hpz} to which we will
refer in the following. We denote with $\sigma\;({\bar\sigma})$
and $\lambda\;({\bar\lambda})$ the helicities of the electron (positron)
and of the $W^-$ $(W^+)$, with $\Theta$ the scattering angle of the 
$W^-$ with respect to the $e^-$ direction in the $e^+ e^-$ center
of mass frame. The polarization amplitude reads:
\be
{\cal M}_{\sigma{\bar\sigma};\lambda{\bar\lambda}}(\Theta)=
\sqrt{2} e^2 {\tilde{\cal M}}_{\sigma{\bar\sigma};\lambda{\bar\lambda}}
(\Theta)~
\epsilon~ {d^{J_0}}_{\Delta\sigma,\Delta\lambda}(\Theta)
\label{a1}
\ee
where $\epsilon=\Delta\sigma (-1)^{\bar\lambda}$, $\Delta\sigma=
(\sigma-\bar\sigma)/2$, $\Delta\lambda=\lambda-\bar\lambda$,
$J_0=max(\vert\Delta\sigma\vert,\vert\Delta\lambda\vert)$ and 
${d^{J_0}}_{\Delta\sigma,\Delta\lambda}$ are the $d$ functions
\cite{ros}.

When $\Delta\lambda=\pm 2$, the reduced amplitude ${\tilde {\cal M}}$ 
contains only the contribution from the neutrino exchange and is given by:
\be
{\tilde{\cal M}}=-\frac{\sqrt{2}}{\sin^2\theta}
\delta_{\Delta\sigma,-1} 
\frac{1}{1+\beta^2-2 \beta \cos\Theta}
\label{a2}
\ee
Here $\beta=\sqrt{1-4 m_W^2/s}$ is the $W$ velocity and $\theta$ is the
weak mixing angle.
For $\vert\Delta\lambda\vert\le 1$, the amplitude is a sum of three 
contributions:
\be
{\tilde{\cal M}}=\Mg+\Mz+\Mnu
\label{a3}
\ee
where
\bea
\Mg &=&-\beta \delta_{|\Delta\sigma|,1} 
\left[\Ag +\dAg \right]\nn\\
\Mz &=&\beta \frac{s}{s-m_Z^2} 
\left[\delta_{|\Delta\sigma|,1}-\frac{\delta_{\Delta\sigma,-1}}{2\sin^2\theta
}\right]
     \left[\Az +\dAz\right]\nn\\
\Mnu &=&\frac{\delta_{\Delta\sigma,-1}}{2\sin^2\theta~\beta}
        \left[B_{\lambda{\bar\lambda}}-
        \frac{1}{1+\beta^2-2 \beta \cos\Theta} C_{\lambda{\bar \lambda}}\right]
\label{a4}
\eea
The coefficients $A$, $\delta A$, $B$ and $C$ contain the dependence on the 
TGV. The terms $A$, $B$ and $C$ represent
the SM, tree-level contribution and they are explicitly listed in Table 1.
%\begin{table}[t]
%\centering
\begin{center}
\begin{tabular}{||c|c|c|c||} \hline
& & & \\
$\lambda\bar\lambda$ & $\Ag = \Az$ & $B_{\lambda \bar\lambda}$ & 
                               $C_{\lambda \bar\lambda}$ \\
& & & \\ \hline
$++ \, ,\,-- $ & $ 1 $ & 1 & $1/ \gamma^2$ \\
$+0 \, ,\,0-$ & $ 2 \gamma $ & $2\gamma$ & $2(1+\beta)/\gamma$ \\
$0+ \, ,\,-0$ & $2 \gamma $ & $2\gamma$ & $2(1-\beta)/\gamma$ \\
$00$ & $2 \gamma^2 + 1$ & $2\gamma^2$ & ${2/ \gamma^2}$ \\ \hline
\end{tabular}
\end{center}
%\caption{ 
%\footnotesize 
\begin{quotation}
Table 1: Standard Model coefficients expressed in terms of 
$\gamma^2=s/4 m_W^2$.
\end{quotation}
%\end{table}
On the contrary, any additional contribution is given by 
$\delta A$ which, assuming exact CP invariance, 
can be decomposed in terms of four form factors:
\bea
\delta A_{++}^V&=&\delta A_{--}^V=\delta f_1^V\nn\\
\delta A_{+0}^V&=&\delta A_{0-}^V=\gamma(\delta f_3^V+\beta\delta f_5^V)\nn\\
\delta A_{-0}^V&=&\delta A_{0+}^V=\gamma(\delta f_3^V-\beta\delta f_5^V)\nn\\
\delta A_{00}^V&=&\gamma^2\left[-(1+\beta^2) \delta f_1^V + 4 \gamma^2 \beta^2 
\delta f_2^V + 2 \delta f_3^V\right]
\label{a5}
\eea                             
The form factors $\delta f^V_i\;(i=1,2,3,5)$ come from the parametrization
of the most general $VWW$ vertex $(V=\gamma,Z)$, compatible with 
Lorentz and $CP$ symmetries and with off-shell components projected out:
\bea
\Gamma^V_{\mu\alpha\beta}(p,q,{\bar q})&=&(f_1^V+\delta f_1^V) 
k_\mu g_{\alpha\beta}-
\delta f_2^V k_\mu \frac{p_\alpha p_\beta}{p^2}\nn\\
& &+(f_3^V+\delta f_3^V) 
(p_\alpha g_{\mu\beta}-
p_\beta g_{\mu\alpha}) + i \delta f_5^V \epsilon_{\mu\alpha\beta\rho} k^\rho
\label{a6}
\eea
where 
$(p,q,{\bar q})$ and $(\mu,\alpha,\beta)$ are the momenta and Lorentz indices
of the $(V,W^-,W^+)$ lines,
$k_\mu=q_\mu-{\bar q}_\mu$
and we have explicitly separated in $f_{1,3}^V$ the SM, tree-level 
contribution: 
\be
f^V_1=1,~~~~f^V_3=2
\label{a8}
\ee
Notice that we have also redefined the form factor $f_2$ of ref. \cite{hpz}
according to:
\be
4\gamma^2 f^{V}_2 \to \delta f^{V}_2 
\label{a9}
\ee
We recall that a factor $-e$ $(-e \cot\theta)$ is conventionally extracted
from the $WW\gamma$ $(WWZ)$ vertex.
The reduced amplitudes of eqs. (\ref{a2},\ref{a4}) allow to analyse
quite clearly the high-energy behaviour of the process.
Referring to the SM, one immediately sees that, due to
the dependence on $\gamma$ of the $A$ and $B$ coefficients,
the separate amplitudes of eqs. (\ref{a4}) with one or both 
$W$'s longitudinally polarized diverge in the large energy limit.
In the sum of eq. (\ref{a3}), those terms cancel as a result 
of the asymptotic equalities:
\be
\Ag=\Az=B_{\lambda{\bar\lambda}}
\label{a10}
\ee
as required by unitarity
\footnote{Moreover, tree-level unitarity constraints can be 
obtained on combinations of $\delta f^V_i$\cite{bau}.} 
.

All the above formulae and properties are standard and well known 
and we have reported them here only to keep our presentation 
self-contained. 

We also recall that, when discussing non standard contributions to 
$e^+ e^- \to W^+ W^-$, it is current practice to assume that
any deviation from the SM predictions is concentrated in
the TGV. 
Departures from the SM at the level
of vector boson self-energies or in $e^+ e^- V~~~~(V=\gamma,Z)$ vertices 
are assumed to be negligibly small in comparison to the deviations 
which can occur in TGV. This assumption is certainly well supported
by the overall amount of data accumulated by LEP1 and SLC,
which have made possible to test vector boson two-point functions
and fermion-antifermion-gauge boson vertices at the per mille level.
However, in any SM extension one can think of, it is quite
difficult, barring unnatural cancellation, to satisfy 
the previous assumption. Deviations from the SM naturally occur
in two- as well as three-point functions for the electroweak bosons.
Indeed, this fact has been exploited to point out that it is 
implausible to expect large deviations from the SM in W pair 
production at LEP2, since the modifications of the TGV required to produce an
observable effect would have probably had a visible counterpart at LEP1/SLC
\cite{der}.
In the present note 
%we will relax the above mentioned assumption.
%and we will accordingly modify the previous formalism.
%In particular, 
we will assume that the underlying theory possesses the following
properties:

(i) Gauge invariance under $SU(2)_L\otimes U(1)_Y$ and discrete CP invariance.

(ii) The deviations from the SM induced by
the new particles or interactions are all of oblique 
type, that is they only affects the $n$-point gauge 
vector boson functions $(n=2,3,4,...)$.

(iii) The deviations from the SM are small and they can be
treated perturbatively in some parameter. In practice we
will discuss W pair production in the one-loop approximation. 

To specify two- and three-point functions for the gauge vector bosons
we will perform the computation in the framework of the background field
gauge, by choosing the t'Hooft-Feynman gauge for the quantum fields.
Then, in any theory fulfilling the properties (i)-(iii), at one-loop level 
the amplitude for $e^+ e^- \to W^+ W^-$ can be cast in the following form
\cite{cul}:
\vskip .5cm

\noindent
$\bullet \Delta\lambda=\pm2$
\be
{\tilde{\cal M}}=-\frac{\sqrt{2}}{\sin^2\thetab}
\delta_{\Delta\sigma,-1} \left[1-\frac{\sin^2\thetab}{\cos 2\thetab} 
\Delta r_W-e_6\right]
\frac{1}{1+\beta^2-2 \beta \cos\Theta}
\label{a11}
\ee

\noindent
$\bullet \Delta\lambda=\le 1$
\bea
\Mg &=&-\beta \delta_{|\Delta\sigma|,1} \left[1+\Delta\alpha(s)\right]
\left[\Ag +\dAg (s)\right]\nn\\
\Mz &=&\beta \frac{s}{s-m_Z^2} 
\left[\delta_{|\Delta\sigma|,1}-\frac{\delta_{\Delta\sigma,-1}}{2\sin^2\thetab
(1+\Delta k(s))}\right]\nn \\
   & &~~~~~~~~~~~~~~~~~~~~~~~~\left[1+\Delta\rho(s)+
     \frac{\cos2\thetab}{\cos^2\thetab}\Delta k(s)\right]
     \left[\Az +\dAz (s)\right]\nn\\
\Mnu &=&\frac{\delta_{\Delta\sigma,-1}}{2\sin^2\thetab~\beta}
        \left[1-\frac{\sin^2\thetab}{\cos 2\thetab} 
        \Delta r_W-e_6\right]\left[B_{\lambda{\bar\lambda}}-
        \frac{1}{1+\beta^2-2 \beta \cos\Theta} C_{\lambda{\bar \lambda}}\right]
\label{a12}
\eea

Here the coefficients $\dAV$ are still given by eq. (\ref{a5}) 
where the form factors $\delta f_i^V (s)~~~(i=1,2,3,5)$,
which in our convention include only the contribution coming 
from the unrenormalized, irreducible, 1-loop correction to the 
vertex $VWW$, should be replaced by combinations $\delta F_i^V(s)$ which 
account for the wave function renormalization of the external 
$W$ legs. This replacement is explicitly given by:
\bea
\delta F_1^V (s) &=& \delta f_1^V (s) - \Pi'_{WW}(m_W^2)\nn\\
\delta F_2^V (s) &=& \delta f_2^V (s)\nn\\
\delta F_3^V (s) &=& \delta f_3^V (s) - 2 \Pi'_{WW}(m_W^2)\nn\\
\delta F_5 (s) &=& \delta f_5^V (s)
\label{a12bis}
\eea
Whereas the form factors $\delta f_{1,3}^V(s)$ are ultraviolet divergent, 
the quantities $\delta F_i^V(s)$ are all finite. 
%Other choices in the renormalization's conditions are equivalent 
%(see for example \cite{peskin}).

The quantities $\Delta\alpha(s)$, $\Delta k(s)$, $\Delta\rho(s)$,  
$\Delta r_W$ and $e_6$ appearing in eqs. (\ref{a11}) and (\ref{a12})
are finite self-energy corrections, defined by: 
\bea
\Delta\alpha(s)&=&\Pi'_{\gamma\gamma}(0)-\Pi'_{\gamma\gamma}(s)\nn\\
\Delta k(s)&=&-\frac{\cos^2\thetab}{\cos 2\thetab}~ (e_1-e_4)+
              \frac{1}{\cos 2\thetab}~ e_3(s)\nn\\
\Delta\rho(s)&=&e_1-e_5(s)\nn\\
\Delta r_W&=&-\frac{\cos^2\thetab}{\sin^2\thetab}~e_1+
\frac{\cos 2\thetab}{\sin^2\thetab}~e_2 + 2~e_3(m^2_Z) + e_4\nn\\ 
e_6&=&\Pi'_{WW}(m_W^2)-\Pi'_{WW}(0)
\label{a13}
\eea
where
\bea
e_1&=&\frac{\Pi_{ZZ}(0)}{m_Z^2}-\frac{\Pi_{WW}(0)}{m_W^2}\nn\\
e_2&=&\Pi'_{WW}(0)-\cos^2\thetab~
\Pi'_{ZZ}(0)-2\cos\thetab\sin\thetab~
\frac{\Pi_{\gamma Z}(m_Z^2)}{m_Z^2}\nn\\
 & & - \sin^2\theta~\Pi'_{\gamma\gamma}(m_Z^2)\nn\\
e_3(s)&=&\frac{\cos\thetab}{\sin\thetab}\left\{\sin\thetab\cos\thetab\left[
\Pi'_{\gamma\gamma}(m_Z^2)-\Pi'_{ZZ}(0)\right]+
\cos 2\thetab~ \frac{\Pi_{\gamma Z}(s)}{s}\right\}\nn\\
e_4&=&\Pi'_{\gamma\gamma}(0)-\Pi'_{\gamma\gamma}(m_Z^2)\nn\\
e_5(s)&=&\Pi'_{ZZ}(s)-\Pi'_{ZZ}(0)
\label{a14}
\eea
In the previous eqs. $\Pi_{ij}~~~(i,j=\gamma,Z~{\rm or}~ W)$ stands for the 
transverse part of the unrenormalized gauge vector bosons self-energy and
\bea
\Pi'_{VV'}(s)=&\dd\frac{\Pi_{VV'}(s)-\Pi_{VV}(m_{VV'}^2)}{(s-m_{VV'}^2)}
~~~~~~~~~~~(V,V'=\gamma,Z,W)
\label{a15} 
\eea
with $m_{\gamma \gamma}=m_{\gamma Z}=0$, $m_{ZZ}=m_Z$ and $m_{WW}=m_W$.

Finally, the effective weak angle $\thetab$ is defined by:
\be
\sin^2\thetab=\frac{1}{2}-\sqrt{\frac{1}{4}-\frac{\pi\alpha(s)}{\sqrt{2} 
G_F m_Z^2}}
\label{a16} 
\ee
where $\alpha(s)$ is the electromagnetic coupling with all the effects 
coming from SM particles included at the given energy $s$.  

Some comments are in order. First of all we will neglect
SM radiative corrections \cite{vel}
in this analysis. They constitute
a gauge-invariant set of corrections and the partial amplitudes they give rise
satisfy unitarity by themselves. Here we would like to focus 
on the contribution coming from new physics.  
We have made use of an on-shell renormalization 
scheme based on the renormalized parameters $\alpha$, $G_F$ and $m_Z$. The 
function $\Delta\alpha(s)$ describes the running of $\alpha$ due to the new
particles; $\Delta k(s)$ represents an energy dependent shift of
the weak effective angle; $\Delta\rho(s)$ is related to the different strengths
of the neutral and charged current interactions; $e_6$ is due to the 
wave-function renormalization of the external $W$'s in the amplitude with
neutrino exchange and $\Delta r_W$ arises when expressing the 
electron-neutrino-W coupling in terms of $\alpha$ and $\sin^2\thetab$.
If $s=m_Z^2$ then $\Delta\alpha(s)$, $\Delta k(s)$, $\Delta\rho(s)$
coincides with the corrections $\Delta\alpha$, $\Delta k$, $\Delta\rho$
which characterize the electroweak observables at the $Z$ resonance \cite{eps,
bfc,stu}.
%Appropriate combinations of such corrections define the so-called 
%$\epsilon$ parameters:
%\bea
%\epsilon_1&=&e_1-e_5(m_Z^2)\nn\\
%\epsilon_2&=&e_2-\sin^2\thetab~ e_4 -\cos^2\thetab ~e_5(m_Z^2)\nn\\
%\epsilon_3&=&e_3(m_Z^2)+\cos^2\thetab~ e_4-\cos^2\thetab~ e_5(m_Z^2)
%\label{a17} 
%\eea

We now consider the high-energy limit of the above amplitudes.
When the energy is much larger than the characteristic scale $M$
of the considered theory, the tree-level asymptotic relations
$\Ag=\Az=B_{\lambda{\bar\lambda}}$ are not sufficient to guarantee
the correct behaviour for the amplitude with one or both longitudinally
polarized $W$'s.
With the inclusion of one-loop contributions, one has new terms 
proportional to $\gamma^2$ and $\gamma$ (see $\dAg$ and $\dAz$ in eq.
(\ref{a12})) and the cancellation of those terms in the high-energy limit 
entails relations among oblique and vertex corrections.
By requiring the asymptotic cancellation of the terms proportional 
to positive powers of $\gamma$, separately for the $\Delta\sigma=1$
and the $\Delta\sigma=-1$ amplitudes, we obtain:
\bea
\left[(\Delta\alpha(s)-\Delta\rho(s)-\frac{\cos2\theta}{\cos^2\theta}
\Delta k(s)) \Ag + \dAg-\dAz\right]_{\infty}&=&{\rm constant}\nn\\
\left[(\Delta\rho(s)-\frac{\sin^2\theta}{\cos^2\theta}
\Delta k(s)) +\frac{\sin^2\theta}{\cos 2\theta} \Delta r_W +e_6)\Ag + \dAz
\right]_{\infty}&=&{\rm constant}
\label{a18}
\eea
where the suffix $\infty$ indicates that the relations hold only for
asymptotically large energies. In eq. (\ref{a18}) at least one 
of the helicities $\lambda$, ${\bar\lambda}$ is required to vanish;
when CP invariance is assumed, this occurs for three independent 
helicity combinations $(\lambda {\bar\lambda})$. This makes a total 
of six independent relations, hereafter termed sum rules. These relations
can be also expressed in terms of unrenormalized self-energies and TGV. 
Using eqs. (\ref{a5}), (\ref{a12bis}), (\ref{a13}), (\ref{a14}), (\ref{a15}), it is 
possible to write the sum rules in eq. (\ref{a18}) in the equivalent form:
\bea
\left[2 \delta f_1^\gamma(s)-\delta f_2^\gamma(s)\right]_\infty &=&
\left[\delta f_3^\gamma(s)\right]_\infty=
\frac{2}{s}\left[\Pi_{\gamma\gamma}(s) +\frac{\cos\theta}{\sin\theta} 
\Pi_{\gamma Z}(s)\right]_\infty\nn\\
\left[\delta f_5^\gamma(s)\right]_\infty &=&0 \nn\\
\left[2 \delta f_1^Z(s)-\delta f_2^Z(s)\right]_\infty &=&
\left[\delta f_3^Z(s)\right]_\infty=
\frac{2}{s}\left[\Pi_{ZZ}(s) +\frac{\sin\theta}{\cos\theta} 
\Pi_{\gamma Z}(s)\right]_\infty\nn\\
\left[\delta f_5^Z(s)\right]_\infty &=&0 
\label{a19}
\eea
Before illustrating
how, in specific models, these sum rules are satisfied, we
will discuss how they are related to the symmetry properties
of the underlying theory.
\vskip 0.5cm

Green functions of spontaneously broken gauge theories satisfy 
Ward Identities (WI), which take a particularly simple form in the
background field gauge formalism \cite{abb}. The case of $SU(2)_L\otimes U(1)_Y$
has been explicitly worked out in ref. \cite{den1}
\footnote{Analogous WI can be derived in one-loop approximation in the
framework of the pinch-technique \cite{pward2}.}. 
Relevant to our discussion
are the WI relating TGV, gauge vector boson self-energies and
vertex functions with two gauge vector bosons and a goldstone boson
$\varphi$:
\be
q^\alpha \Gamma^{V W W}_{\mu\alpha\beta}(p,q,{\bar q})+
i m_W \Gamma^{V \varphi W}_{\mu\beta}(p,q,{\bar q})=
\left[\Pi^V_{\mu\beta}(p)-\Pi^{WW}_{\mu \beta}({\bar q})\right]\nn\\
\label{a22}
\ee
where $V=\gamma,Z$ and the functions $\Pi^V_{\mu\beta}(p)$ are defined by:
\bea
\Pi^\gamma_{\mu\beta}(p)&=&\Pi^{\gamma\gamma}_{\mu\beta}(p)+\frac{\cos\theta}
{\sin\theta}\Pi^{\gamma Z}_{\mu \beta}(p)\nn\\
\Pi^Z_{\mu\beta}(p)&=&\Pi^{Z Z}_{\mu\beta}(p)+\frac{\sin\theta}{\cos\theta}
\Pi^{\gamma Z}_{\mu \beta}(p)
\label{a22b}
\eea
These identities are direct consequence of the $SU(2)_L\otimes U(1)$
gauge invariance and we are naturally lead to explore the relation
between the sum rules of eq. (\ref{a19}) and the WI listed above.
Indeed, as we will now show, the sum rules discussed here can be
directly derived from the WI of the theory. To this purpose we
consider the general, off-shell, decomposition of the $VWW$
and $V\varphi W$ vertices:
\bea
\Gamma^{VWW}_{\mu\alpha\beta}(p,q,{\bar q})&=&\delta {\bar f}_1^V 
k_\mu g_{\alpha\beta}-
\delta {\bar f}_2^V k_\mu \frac{p_\alpha p_\beta}{p^2}+
\delta {\bar f}_3^V (p_\alpha g_{\mu\beta}-
p_\beta g_{\mu\alpha}) +\nn\\
& & i \delta {\bar f}_5^V 
\epsilon_{\mu\alpha\beta\rho} k^\rho
-\delta h_2^V\frac{k_\alpha k_\beta k_\mu}{p^2} 
+\delta h_3^V(k_\alpha g_{\mu\beta}+k_\beta g_{\mu\alpha})\nn\\
& &
-\delta h_4^V(k_\alpha p_\beta-k_\beta p_\alpha)\frac{k_\mu}{p^2}
-\delta h_5^V(k_\alpha p_\beta+k_\beta p_\alpha)\frac{p_\mu}{p^2}\nn\\
& &
+i \delta k_5^V \epsilon_{\mu\nu\rho\sigma} \dd\frac{p^\rho k^\sigma}{p^2}
[(p+k)_\alpha \delta_\beta^\nu-(p-k)_\beta \delta_\alpha^\nu]
+ i \delta k_6^V\epsilon_{\alpha\beta\rho\sigma}\frac{p^\rho 
k^\sigma}{p^2} p_\mu
\label{a23}
\eea
\bea
i \Gamma^{V\varphi W}_{\mu\beta}(p,q,{\bar q})&=&
\delta\varphi_1^V g_{\mu \beta}+
\delta\varphi_2^V \dd\frac{p_\mu p_\beta}{p^2}+
\delta\varphi_3^V \dd\frac{{\bar q}_\mu p_\beta}{p^2}\nn\\
& &+
\delta\varphi_4^V \dd\frac{p_\mu {\bar q}_\beta}{p^2}+
\delta\varphi_5^V \dd\frac{{\bar q}_\mu {\bar q}_\beta}{p^2}+
i \delta\varphi_6^V 
\epsilon_{\mu\beta\rho\sigma}\dd\frac{{\bar q}^\rho p^\sigma}{p^2}
\label{a24}
\eea
We also make use of the standard parametrization for the vector 
boson self-energies:
\be
\Pi_{\mu\nu}^{ij}(p)=(g_{\mu\nu}-\dd\frac{p_\mu p_\nu}{p^2}) \Pi^{ij}(p^2)+
\dd\frac{p_\mu p_\nu}{p^2}\Pi_L^{ij}(p^2)
\label{b2}
\ee
Notice that all the form factors of eq. (\ref{a23}) 
are dimensionless function of $p^2$, $q^2$
and ${\bar q}^2$. 
Moreover, when the terms proportional to $p_\mu$,
$q_\alpha$ and ${\bar q}_\beta$ are neglected, as for the case
of the on-shell amplitude of interest, the above decomposition
collapses on that given in eq. (\ref{a6}),
provided one makes the following identifications:
\bea
\delta f_1^V&=&\delta {\bar f}_1^V\nn\\
\delta f_2^V&=&\delta {\bar f}_2^V-\delta h_2^V-2 \delta h_4^V\nn\\
\delta f_3^V&=&\delta {\bar f}_3^V-\delta h_3^V\nn\\
\delta f_5^V&=&\delta {\bar f}_5^V
\label{b0}
\eea
%On the other hand, the form factors $\delta\varphi_i^V~~~(i=1,...6)$
%have the dimension of a mass. 
%When contracted with $q_\alpha$ the trilinear vertex $VWW$ gives:
%\bea
%q^\alpha \Gamma^{VWW}_{\alpha\beta\mu}&=&\dd\frac{1}{2}\left[
%p^2 g_{\mu\beta} \delta f_3^V
%+p_\mu p_\beta \left(2 \delta f_1^V -\delta f_2^V -2\delta f_3^V\right)
%+{\bar q}_\mu p_\beta \left(-4 \delta f_1^V +2\delta f_2^V +2\delta f_3^V
%\right)\right.\nn\\
%& &\left.+p_\mu {\bar q}_\beta \left(-2 \delta f_1^V -2\delta h_2^V 
%-4\delta h_3^V
%-2 \delta h_4^V +2\delta h_5^V\right)\right.\nn\\
%& &\left.
%+{\bar q}_\mu {\bar q}_\beta \left(4 \delta f_1^V +4\delta h_2^V +
%4\delta h_3^V+4 \delta h_4^V\right)\right.\nn\\
%& &\left. -2i\epsilon_{\mu\beta\rho\sigma} {\bar q}^\rho p^\sigma
%\left(\delta f_5^V-4 \delta k_5^V\dd\frac{m_W^2}{p^2}\right)\right]
%\label{b1}
%\eea
When we combine eqs. (\ref{a23}), (\ref{a24}) and (\ref{b2}) with
eqs. (\ref{a22}), after identifications of the independent
tensor structures, we obtain the following relations:
\bea
\dd\frac{p^2+pk}{2}\delta {\bar f}_3^V+\dd\frac{pk+k^2}{2}\delta h_3^V+
m_W \delta\varphi_1^V&=&\Pi^V(p^2)-\Pi^{WW}({\bar q}^2)\nn\\
\delta {\bar f}_1^V-\dd\frac{p^2+pk}{2 p^2} \delta {\bar f}_2^V
-\dd\frac{pk+k^2}{2 p^2} \delta h_2^V 
+\dd\frac{p^2-k^2}{2 p^2} \delta h_4^V - ~~~~~~~~~~& &\nn\\
- \delta{\bar f}_3^V+\delta h_3^V
-\dd\frac{p^2+2pk+k^2}{2 p^2} \delta h_5^V 
+\dd\frac{m_W}{p^2} \delta\varphi_2^V&=&\dd\frac{\Pi_L^V(p^2)}{p^2}
-\dd\frac{\Pi^V(p^2)}{p^2}\nn\\
-2 \delta {\bar f}_1^V
+\dd\frac{p^2+pk}{p^2}\delta {\bar f}_2^V 
+\dd\frac{pk+k^2}{p^2}\delta h_2^V
-\frac{p^2-k^2}{p^2} \delta h_4^V + ~~~~~~~~~~& &\nn\\
+ \delta {\bar f}_3^V-\delta h_3^V+\dd\frac{m_W}{p^2} \delta\varphi_3^V
&=&0\nn\\
\dd\frac{p^2+2pk+k^2}{p^2} \delta k_5^V-\delta {\bar f}_5^V
+\dd\frac{m_W}{p^2}\delta\varphi_6^V&=&0\\
-\delta {\bar f}_1^V+\dd\frac{pk+k^2}{p^2}\delta h_2^V
-2 \delta h_3^V- ~~~~~~~~~~~~~~~~~~~~~~~~~~~~~~& &\nn\\
- \dd\frac{p^2+pk}{p^2}\delta h_4^V
+\dd\frac{p^2+pk}{p^2} \delta h_5^V
+\dd\frac{m_W}{p^2}\delta\varphi_4^V&=&0\nn\\
2 \delta {\bar f}_1^V -2\dd\frac{pk+k^2}{p^2}\delta h_2^V
+2 \delta h_3^V + ~~~~~~~~~~~~~~~~~~~~~~~~~~~~~~& &\nn\\
+ 2\dd\frac{p^2+pk}{p^2} \delta h_4^V
+\dd\frac{m_W}{p^2}\delta\varphi_5^V&=&
\dd\frac{\Pi^{WW}({\bar q}^2)}{{\bar q}^2}
-\dd\frac{\Pi_L^{WW}({\bar q}^2)}{{\bar q}^2}\nn\\
& &\nn
\label{b3}
\eea
%\bea
%\dd\frac{p^2}{2} \delta f_3^\gamma+m_W \delta\varphi_1^\gamma&=&
%\Pi^{\gamma\gamma}+\dd\frac{\cos\theta}{\sin\theta}\Pi^{\gamma Z}
%-\Pi^{WW}({\bar q}^2)\nn\\
%\delta f_1^\gamma-\dd\frac{1}{2}\delta f_2^\gamma
%-\delta f_3^\gamma+m_W \dd\frac{\delta\varphi_2^\gamma}{p^2}&=&
%\dd\frac{(\Pi_L^{\gamma\gamma}-\Pi^{\gamma\gamma})}{p^2}
%+\dd\frac{\cos\theta}{\sin\theta}
%\dd\frac{(\Pi_L^{\gamma Z}-\Pi^{\gamma Z})}{p^2}\nn\\
%-2\delta f_1^\gamma+\delta f_2^\gamma
%+\delta f_3^\gamma+ m_W \dd\frac{\delta\varphi_3^\gamma}{p^2}&=&0\nn\\
%\dd\frac{p^2}{2} \delta f_3^Z+m_W \delta\varphi_1^Z&=&
%\Pi^{ZZ}+\dd\frac{\sin\theta}{\cos\theta}\Pi^{\gamma Z}
%-\Pi^{WW}({\bar q}^2)\nn\\
%\delta f_1^Z-\dd\frac{1}{2}\delta f_2^Z
%-\delta f_3^Z+m_W \dd\frac{\delta\varphi_2^Z}{p^2}&=&
%\dd\frac{(\Pi_L^{ZZ}-\Pi^{ZZ})}{p^2}
%+\dd\frac{\sin\theta}{\cos\theta}
%\dd\frac{(\Pi_L^{\gamma Z}-\Pi^{\gamma Z})}{p^2}\nn\\
%-2\delta f_1^Z+\delta f_2^Z
%+\delta f_3^Z+m_W \dd\frac{\delta\varphi_3^\gamma}{p^2}&=&0\nn\\
%\delta f_5^{\gamma,Z}-4\delta k_5^{\gamma,Z}\dd\frac{m_W^2}{p^2}-
%\dd\frac{m_W \delta\varphi_6^{\gamma,Z}}
%{p^2}=0\nn
%\label{b3}
%\eea
We now impose the on-shell conditions $q^2={\bar q}^2=m_W^2$ and
we proceed to expand the above
equations in inverse power of $p^2$.
In doing that, we assume that for each form
factor the leading term of this expansion is provided by naive
dimensional analysis. In other words, the dimensionless functions
$\delta {\bar f}_i^V$, $\delta h_i^V$ and $\delta k_i^V$ are assumed 
not to grow with positive powers of $p^2$. The two-point functions 
$\Pi^{ij}$ and $\Pi_L^{ij}$ are required to scale at most like $p^2$
and the form factors $\delta\varphi_i^V$ may scale as $\sqrt{p^2}$.
Logarithmic corrections to this behaviour are also admitted
\footnote{Notice that the factors of $p^2$ inserted in eq. 
(\ref{a23}) to obtain dimensionless form factors are consistent
with this assumption. A similar comment applies to eq. (\ref{a24}).}.
By identifying, order by order, the expanded expressions one
obtains an infinite set of relations. Those
corresponding to the leading terms, after using eq. (\ref{b0}) are 
given by
\footnote{One can check that some of the relations obtained at the
next-to-leading order coincide with those predicted by the so-called
equivalence theorem \cite{clt}.}:
\bea
\left[\delta f_3^V(p^2)\right]_\infty&=&\dd\frac{2}{p^2}
\left[\Pi^V(p^2)\right]_\infty\nn\\
\left[-2\delta f_1^V(p^2)+\delta f_2^V(p^2)+\delta f_3^V(p^2)\right]_\infty 
&=& 0\nn\\
\left[\delta f_1^V(p^2)-\dd\frac{\delta f_2^V(p^2)}{2}-
\delta f_3^V(p^2)\right]_\infty &=&
-\dd\frac{1}{p^2}\left[\Pi^V(p^2)-\Pi_L^V(p^2)\right]_\infty \nn\\
\left[\delta f_5^V(p^2)\right]_\infty &=& 0\\
\left[\delta f_1^V(p^2)+\delta h_2^V(p^2)+2\delta h_3^V(p^2)
+\delta h_4^V(p^2)-\delta h_5^V(p^2)\right]_\infty &=& 0\nn\\
\left[2\delta f_1^V(p^2)+2\delta h_2^V(p^2)+2\delta h_3^V(p^2)
+2\delta h_4^V(p^2)\right]_\infty &=& \dd\frac{1}{{\bar q}^2}
\left[\Pi^{WW}({\bar q}^2)-\Pi_L^{WW}({\bar q}^2)\right]\nn
\label{b3b}
\eea
From these equations one can immediately recover
the sum rules of eq. (\ref{a19}). One also finds $\left[\Pi_L^V(p^2)\right]_\infty=0$
and two conditions on the form factors $\delta h_i^V~~~(i=2,...5)$.
Had we used the less general parametrization for the vertex functions
with $\delta h_i^V=0~~~(i=2,...5)$, we would have obtained inconsistent
equations. We notice that, in the derivation which make use of the
WI of eq. (19), the assumption (ii) is inessential.
We are thus led to conjecture that in the presence of corrections 
involving the electron-positron lines, such corrections should
satisfy a set of independent asymptotic relations which do not
interfere with those obtained here. 
 
\vspace{1cm}
{\bf 3.} Eq. (\ref{a19}), or its equivalent form eq. (\ref{a18}), represents
the main result of the present work. Their meaning can be better elucidated
with some examples.

For instance, the 1-loop contribution due to an additional, heavy quark doublet
with a degenerate mass $M$ reads:
\bea
\delta f_1^\gamma(s)&=&\dd\frac{g^2}{16\pi^2}\left[ A + 
\dd\frac{7}{6}
\right]+...\nn\\
\delta f_2^\gamma(s)&=&-\dd\frac{g^2}{16\pi^2}
+...\nn\\
\delta f_3^\gamma(s)&=&\dd\frac{g^2}{16\pi^2}\left[2 A + 
\dd\frac{10}{3}
\right]+...\nn\\
\delta f_5^\gamma(s)&=&0\nn\\
\Pi_{\gamma\gamma}(s)&=&\dd\frac{g^2}{16\pi^2}\sin^2\theta\left[\dd\frac{20}
{9} A +
\dd\frac{100}{27}\right]s+...\nn\\
\Pi_{\gamma Z}(s)&=&\dd\frac{g^2}{16\pi^2}\dd\frac{\sin\theta(9-20\sin^2\theta)}
{\cos\theta}\left[\dd\frac{1}{9} A +\dd\frac{5}{27}\right]s+...\nn\\
\Pi_{Z Z}(s)&=&\dd\frac{g^2}{16\pi^2}\dd\frac{(9-18\sin^2\theta+20\sin^4\theta)}
{\cos^2\theta}\left[\dd\frac{1}{9} A +\dd\frac{5}{27}\right]s+...
\label{a20}
\eea
where 
\be
A=\frac{2}{d-4}+ i\pi-{\rm Log}\left(\frac{s}{\mu^2}\right)
\label{a21}
\ee
and dots stand for terms of order $M^2/s$ or $m_W^2/s$. 
The leading terms of $\delta f_i^Z(s)$ in the large $s$ limit
coincide with those of $\delta f_i^\gamma (s)$ given above.
The divergence contained in the expression $A$ is cancelled in
the combinations $\Delta\alpha(s),~~\Delta k(s),~~\Delta\rho(s)$,
$\Delta r_W$ and $e_6$
or when including the $W$ wave function renormalization in the $VWW$
vertices, as explicitly shown in eq. (\ref{a12bis}).

The asymptotic expressions of the previous equations satisfy the sum rules
given in eq. (\ref{a18}).
The form factors $\delta f_{1,2,3}^V(s)$ do not vanish 
at large energies, not even when combined with the $W$ wave function
renormalization. Rather, the correct high-energy 
behaviour of the $e^+ e^- \to W^+ W^-$ amplitudes is assured by the 
interplay between vector boson self-energies and TGV. 
For a new heavy quark doublet, we have explicitly verified that the 
cancellation implied by eq. (\ref{a18}) holds only asymptotically.
In this example, at lower energies, 
the relations in eq. (\ref{a18}) are corrected by terms of order $M^2/s$.
Indeed no unitarity argument forbids such terms, which, enhanced by
the additional longitudinal factors $\gamma$ or $\gamma^2$, may lead to 
large and observable deviations from the SM amplitude, a behaviour
known as delayed unitarity \cite{peskin}.

As a second example we consider the effect of heavy electroweak gauginos, with
a common mass $M$, in the minimal supersymmetric standard model. 
We assume that squarks, sleptons, higgsinos and additional higgses
decouple from the low-energy theory due to their masses which we take 
much larger than $M$. In this case, we find it useful to give
the complete result, up to terms of order $m_W^2/s$:
\bea
\delta f_1^{\gamma,Z}(s)&=&\dd\frac{g^2}{16\pi^2}
\left[\dd\frac{4}{3}\left(A'
-2\dd\frac{s-4 M^2}{s} B \right)
-\dd\frac{2}{9 s} \left(48 M^2 -7 s -36 M^2 C\right)
\right]\nn\\
\delta f_2^{\gamma,Z}(s)&=&\dd\frac{g^2}{16\pi^2}
\left[32\dd\frac{M^2}{s} B -
\dd\frac{4}{3 s}\left(24 M^2 + s -12 M^2 C
\right)\right]\nn\\
\delta f_3^{\gamma,Z}(s)&=&\dd\frac{g^2}{16\pi^2}
\left[\dd\frac{8}{3}\left(A'
-2\dd\frac{s+2 M^2}{s} B \right)
+\dd\frac{8}{9 s} \left(12 M^2 +5 s\right)\right]\nn\\
\delta f_5^{\gamma,Z}(s)&=&0\nn\\
\Pi_{ZZ}(s)&=&\dd\frac{g^2}{16\pi^2} \cos^2\theta
\left[\dd\frac{4}{3}\left(A'
-2 (s+2 M^2) B \right) s+
\dd\frac{4}{9}(12 M^2+5 s)\right]\nn\\
&=&\dd\frac{\cos^2\theta}{\sin^2\theta}
\Pi_{\gamma\gamma}(s)
=\dd\frac{\cos\theta}{\sin\theta}
\Pi_{\gamma Z}(s)
\eea
where
\bea
A'&=&\dd\frac{2}{4-d}-{\rm Log}\dd\frac{M^2}{\mu^2}\nn\\
B&=&\sqrt{-1+\dd\frac{4 M^2}{s}}{\rm ArcTan}\dd\frac{1}
{\sqrt{-1+\dd\frac{4 M^2}{s}}}\nn\\
C&=&{\rm ArcTan}\dd\frac{1}
{(-1+\dd\frac{4 M^2}{s})}
\eea
We notice that in this case, the sum rules of eq. (\ref{a18}) are
satisfied not only asymptotically, for $s$ much larger than $M$,
but also at lower energies. Actually, neglecting terms of order
$m_W^2/s$, they are satisfied identically in $s$.
This means that in this case the cancellation dictated by
unitarity take place before the asymptotic regime. The
enhancement factors provided by the longitudinal $W$ components
get now multiplied by terms of order $m_W^2/s$ in the overall amplitude
and deviations from the SM larger than few per cent cannot be expected
\cite{cul}.
Quite a different result would have been obtained if one had
neglected the contribution from the vector boson self-energies.
In this case no unitarity cancellation would have occurred and,
due to the unbalanced $\gamma$ or $\gamma^2$ contributions,
one would have obtained large, unrealistic departures from the 
SM cross-section, even at energies below the threshold for production of new 
particles. This is exemplified in fig. 1 where we compare the
cross-section for longitudinally polarized $W$'s (LL)
obtained, for gauginos, with or without including the neutral gauge 
boson self-energies. 
\begin{figure}[h]
%\vspace{-0.1cm}
\begin{tabular}{cc}
\epsfig{figure=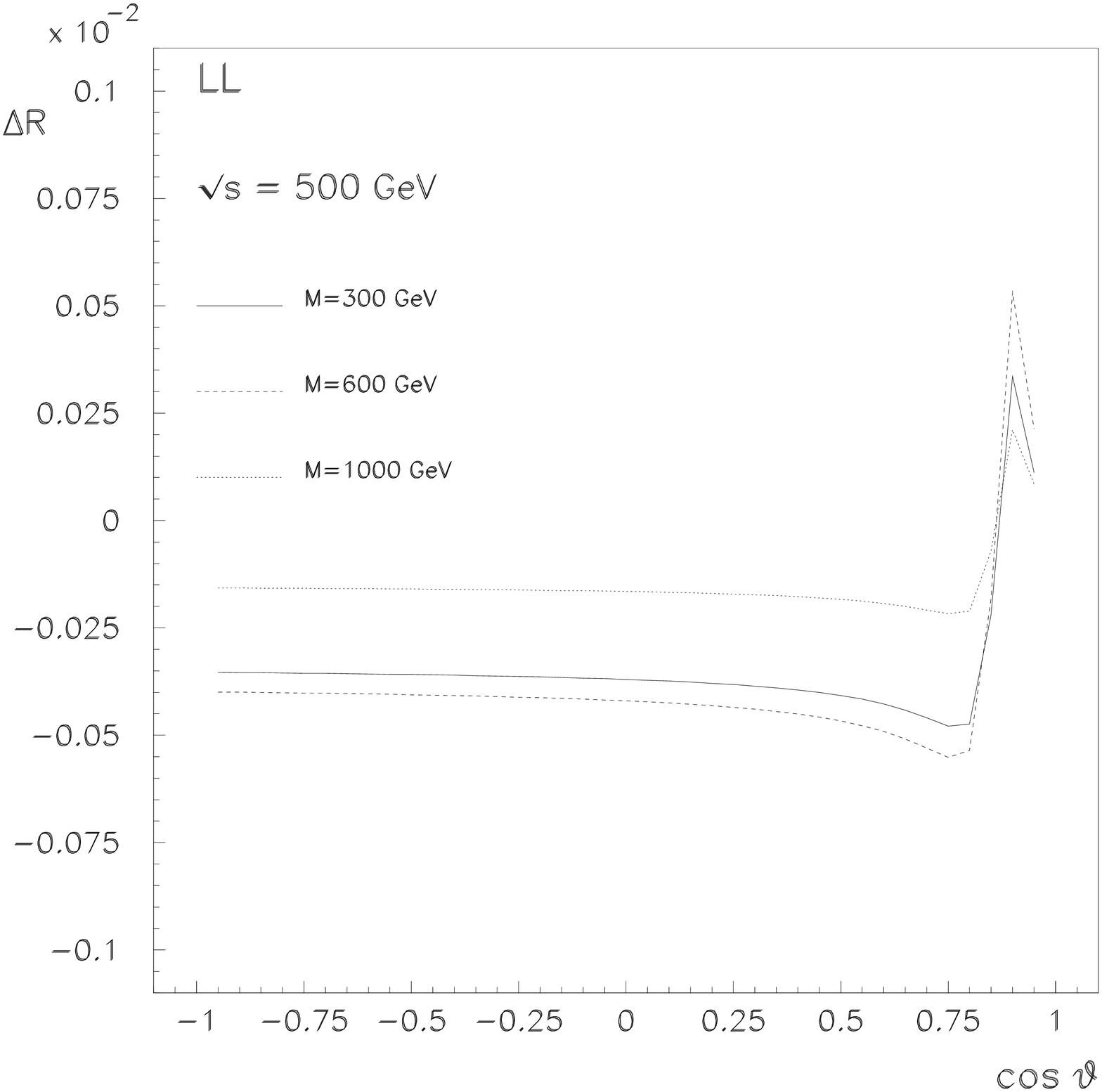,height=7.6cm,angle=0} &
\epsfig{figure=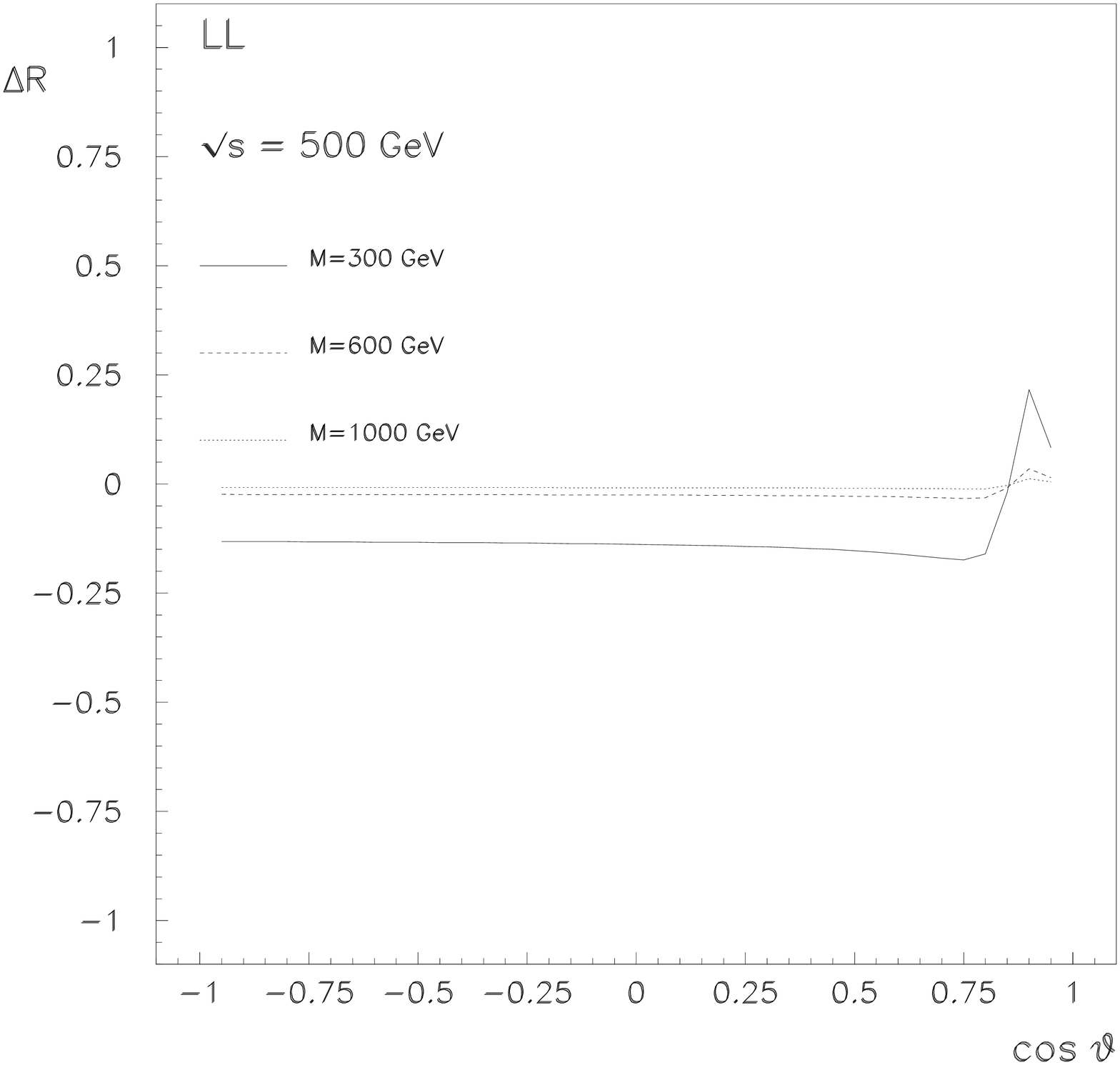,height=7.6cm,angle=0}\\
\end{tabular}
%\vspace{-0.3cm}

\caption{\small \it Relative deviation $\Delta R$ from the SM cross-section 
$d\sigma/d\cos\Theta$ versus $\cos\Theta$ in the LL polarization channel 
at $\sqrt{s}=500~{\rm GeV}$. 
The figure on the left shows the contribution due to gauginos of mass 
respectively $M=300,600,1000$ GeV (full, dotted, dashed) when all the 
contributions (bilinear and trilinear) are included. 
The one on the right is without self-energy contributions.}

\label{true}
\end{figure}
This should sound as a warning against the assumption that
the amplitudes for $e^+ e^- \to W^+ W^-$ are dominated by 
the irreducible TGV corrections. 
\vskip .5cm

{\bf 4.}
The parametrization given in eqs. (\ref{a1}-\ref{a6}) is quite efficient 
and widely used in the literature. Indeed, as will be shown in this section, 
it is possible to cast the results of our one-loop computation in a 
form which is very close to that displayed in eqs. (\ref{a1}-\ref{a6}). 
By inspecting eqs. (\ref{a10}-\ref{a11}), we are led to introduce an
effective, $s$-dependent, Weinberg angle defined by:
\be
\sin^2\theta_{eff}(s)=\sin^2{\bar\theta} (1+\Delta k(s))
\label{a30}
\ee
Then we proceed by absorbing the overall correction to the
neutrino-exchange amplitude in a redefinition of the electric charge:
\be
e_{eff}^2(s)=e^2 (1+\Delta k(s)-\dd\frac{\sin^2{\bar\theta}}{\cos 2{\bar\theta}}
\Delta r_W-e_6)
\label{a31}
\ee
When expressed in terms of $\sin^2\theta_{eff}(s)$ and $e^2_{eff}(s)$, 
the amplitudes of eqs. (\ref{a1},\ref{a11},\ref{a12}) will coincide
with those of eqs. (\ref{a1},\ref{a2},\ref{a4}), provided
one defines new coefficients $\delta A$, which, to avoid confusion,
we will denote by $\Delta A$:
\bea
\Delta A^\gamma_{\lambda{\bar\lambda}}(s)&=&
\delta A^\gamma_{\lambda{\bar\lambda}}(s)+
A^\gamma_{\lambda{\bar\lambda}}~ 
(\dd\frac{\sin^2{\bar\theta}}{\cos 2{\bar\theta}} \Delta r_W+e_6+
\Delta \alpha(s)-\Delta k(s))\nn\\
\Delta A^Z_{\lambda{\bar\lambda}}(s)&=&
\delta A^Z_{\lambda{\bar\lambda}}(s)+
A^Z_{\lambda{\bar\lambda}}~
(\Delta\rho(s)-\dd\frac{\sin^2{\bar\theta}}{\cos^2{\bar\theta}} \Delta k(s)
+\dd\frac{\sin^2{\bar\theta}}{\cos 2{\bar\theta}} \Delta r_W+e_6)
\label{a32}
\eea
If we express the quantities $\Delta\alpha(s)$, $\Delta k(s)$, 
$\Delta\rho(s)$, $\Delta r_W$ and $e_6$ in terms of unrenormalized
self-energy corrections, as in eqs. (\ref{a13}) and (\ref{a14}),
and if we relate the coefficients $\delta A$ to the unrenormalized
vertex corrections, we find that eqs. (\ref{a32}) are equivalent to the 
following definition of form factors:
\bea
\Delta f^V_1(s)&=&\delta f^V_1(s)-\widehat{\Pi}^V(s)\nn\\
\Delta f^V_2(s)&=&\delta f^V_2(s)\nn\\
\Delta f^V_3(s)&=&\delta f^V_3(s)-2 \widehat{\Pi}^V(s)\nn\\
\Delta f^V_5(s)&=&\delta f^V_5(s)
\label{a33}
\eea
where the functions $\widehat{\Pi}^V(s)~~~(V=\gamma,Z)$ are explicitly given by:
\bea
\widehat{\Pi}^\gamma(s)&=&\Pi'_{\gamma\gamma}(s)+
\dd\frac{\cos{\bar\theta}}{\sin{\bar\theta}}\dd\frac{\Pi_{\gamma Z}(s)}{s}\nn\\
\widehat{\Pi}^Z(s)&=&\Pi'_{ZZ}(s)+
\dd\frac{\sin{\bar\theta}}{\cos{\bar\theta}}\dd\frac{\Pi_{\gamma Z}(s)}{s}
\label{a34}\eea
To summarize, by replacing the electric charge $e$, the Weinberg angle
$\theta$ and the form factors $\delta f^V_i(s)$ of eqs. (\ref{a1}-\ref{a6})
with the quantities $e_{eff}(s)$, $\theta_{eff}(s)$, $\Delta f^V_i(s)$
introduced above, one reproduces exactly the one-loop amplitude  
discussed in the present paper. In fact, eqs. (\ref{a33}-\ref{a34})
provide a concise and practical prescription to account for the 
self-energy effects in $e^+e^- \to W^+ W^-$, by including them
in appropriately defined form factors.
Moreover, it is immediate to show that the sum rules discussed before,
when expressed in terms of the form factors $\Delta f^V_i(s)$,
read:
\be
\left[2 \Delta f_1^V(s)-\Delta f_2^V(s)\right]_\infty =
\left[\Delta f_3^V(s)\right]_\infty=
\left[\Delta f_5^V(s)\right]_\infty =0 ~~~.
\label{a35}
\ee
When discussing the low-energy limit of the process,
an effective lagrangian description turns out to be useful.
The departures from the tree-level SM predictions
are described by a set of operators (organized in a 
dimensional or derivative expansion) whose coefficients
can be related to physical observables.
In particular the low-energy limit of the form factors
$\Delta f^V_i$ can be expressed in terms of the 
coefficients $a_i~(i=0,...14)$ characterizing the
so-called electroweak chiral lagrangian \cite{abe,hol,mfr}
\footnote{We follow the conventions of the first paper in ref. \cite{mfr}.}
:
\bea
\Delta f^\gamma_1(0)&=&g^2 (a_1-a_8)\nn\\
\Delta f^\gamma_2(0)&=&0\nn\\
\Delta f^\gamma_3(0)&=&g^2 (a_1-a_8+a_2-a_3-a_9)\nn\\
\Delta f^\gamma_5(0)&=&0\nn\\
\Delta f^Z_1(0)&=&-g^2(a_8+a_{13})-g^2\tan^2{\bar \theta}(a_1+a_{13})
-\dd\frac{g^2}{\cos^2{\bar \theta}} a_3\\
\Delta f^Z_2(0)&=&0\nn\\
\Delta f^Z_3(0)&=&-g^2(a_8+a_3+a_9+a_{13})-g^2\tan^2{\bar \theta}(a_1+a_2+a_{13})
-\dd\frac{g^2}{\cos^2{\bar \theta}} a_3\nn\\
\Delta f^Z_5(0)&=&-\dd\frac{g^2}{\cos^2{\bar\theta}} a_{14}\nn
\eea
We notice that the combinations $\Delta f^V_3(0)-\Delta f^V_1(0)~~~(V=
\gamma,Z)$ depend only on the coefficients $a_2$, $a_3$ and $a_9$
which parametrize the directions that are blind to the
LEP1 precision tests \cite{der}.
\vskip 0.5truecm

{\bf 5.} In conclusion, we have shown that, at asymptotically
large energies, the trilinear gauge boson vertices in 
$e^+ e^- \to W^+ W^-$ amplitudes
obey a set of sum rules connecting them to gauge vector
boson self-energies. We have derived the sum rules
by performing an explicit one-loop computation in a certain
class of theories and by demanding that the resulting amplitudes
satisfy the requirement of perturbative unitarity. We have also
demonstrated that these sum rules are a direct consequence of
the Ward identities of spontaneously broken $SU(2)_L\otimes U(1)_Y$
gauge theories and therefore they are verified in a more general context
than the one considered in the first derivation. 
Our discussion shows that, in general, it is never possible to neglect
vector boson self-energies when computing the form factors
which parametrize the $e^+ e^- \to W^+ W^-$ helicity amplitudes.
The exclusion of the self-energy contributions would lead to
estimates of the effects which are wrong by orders
of magnitude, as we have explicitly discussed in some examples. 
Finally we have shown how the self-energy effects can be suitably
included in the formalism by a simple redefinition of the form factors
which allow to use the existing parametrizations of the considered process.   
\section*{Acknowledgements}
We would like to thank Alessandro Culatti, Giuseppe Degrassi, Antonio Masiero,
Joaquim Matias, Luca Silvestrini, Alessandro Vicini and Fabio Zwirner for many useful discussions and comments.
%
%\newpage
%
%

\end{document}